\magnification 1200
\nopagenumbers
\parskip=10pt
\vskip .7truein
\centerline{\bf
{Vacuum fluctuations for spherical gravitational impulsive waves}}
\vskip 1.0truein
\centerline
{M. Horta\c csu}
\vskip .3truein
\centerline
{Physics Department, T\" UBITAK Marmara Research Centre,} 
\centerline{Research Institute for Basic Sciences, Gebze Turkey}
\vskip .1truein
\centerline
{and}
\vskip .1truein
\centerline
{Physics Department, Faculty of Sciences and Letters,
 ITU, 80626, Maslak, Istanbul, Turkey}
\vskip 1.4truein
\noindent
{\bf{Abstract}}
We propose a method for calculating vacuum fluctuations on the background
of a spherical impulsive gravitational wave which results in a finite
expression for the vacuum expectation value of the stress-energy tensor.
The method is based on first including a cosmological constant   as an
auxiliary constant.  We show that the result for the vacuum expectation value
of the   stress-energy tensor in second-order perturbation theory is finite
if both the cosmological constant and the infrared parameter tend to zero
 at the same rate.

\vfill\eject
\pageno=1
\noindent                                                     
{\bf{1.Introduction}}

It has been known for a long time that plane waves do not give rise to vacuum
fluctuations [1,2].  One may check whether the same result is true for
spherical waves as well.
One possible check is to use the Nutku-Penrose metric [3] for spherical
impulsive gravitational waves and to study whether vacuum fluctuations
arise when a scalar particle is
coupled to the gravitational field through this metric.

We applied this procedure to different holomorphic warp functions [4-8]
and found no finite fluctuations in the Minkowski case.  We must stress the
fact that in these calculations first-order perturbation theory was used.
This method gave a null result for the Minkowski case.  Taking only first
order perturbation theory was, perhaps , the weak point in these calculations.
One conjecture [9] was that the null result would change in second-order
calculations.  Here our results on this problem , performed in second-order
perturbation theory, will be reported.

It is common knowledge that massless particle production is prohibited
by the presence of conformal symmetry [10].  One may argue that this fact
is not relevant here since the Nutku-Penrose metric is not conformally
related to Minkowski space.  This would be true if an exact calculation were
made.  One must note at this point that we are using only approximate methods
to solve this problem.  We are perturbing around the Minkowski space to
calculate the fluctuations in the new space.  This perturbation does not
change the essential properties of the solutions of the Minkowski space,
but just modulates them by multiplying by a factor.  The Minkowski Green
function is just a monomial over distance.  We could not
find a way to regularize it and still retain non-zero terms,
since there are none left when the worst divergence is subtracted.  Our
perturbative calculation just modifies the Minkowski expression by
multiplying it by a finite term.
As a result we could not get any finite result in this case in first-order
perturbation theory for the cases studied.

If we had a mass parameter in the problem, we could get an expansion whose
second or third term could give finite fluctuations.
This was not possible here.  We still hoped  that
the second-order perturbation calculation might give a signal for a possible
way out of this impasse.

One way to introduce a mass parameter without changing the Nutku-Penrose
solution too much is to study a similar case in de Sitter space where one
obtains an impulsive wave solution by a simple manoeuvre [11].  The scalar
curvature of the de Sitter space has dimensions of mass squared.  Here the
calculation in first order gives finite results for all of the different
choices of the warp functions studied [8,12].  If we carry the calculations
to second order, at least for the special choice studied here, we get
infrared divergences.
We conjecture that this behaviour is generic for a wider class of warp
functions, if not for all possible forms, since in first-order calculations
we get the same general behaviour for all the different choices we have
studied.  We verify this conjecture for another warp
function which has shown to be representative for a wider class of warp
functions.

We take these divergences for the signal we are looking for.  The Green
function we
calculate includes terms which are inversely proportionaŸ to the square of
an infrared mass which should bectaken to zero at the end of the calculation.
One natural way to get rid of these two parameters (the curvature of the de
Sitter space and the infrared parameters) is to set them proportional to
each other.
In one sense both terms are introduced for technical reasons.  We went to
de Sitter space to perturb around a space which is different from the
Minkowski in the first place, but one that still retains the Nutku-Penrose
solution.  The infrared parameter was introduced just to be able to calculate
the Green function unambigiously.  There were no massive parameters in the
model initially, and they should not be present in the end result.
Taking these two terms proportional to each other and then sending them to
zero cancels these two auxiliary parameters completely, still retaining a
finite contribution to $<T_{\mu \nu}  >$.

At the end of the calculation all reference to the de Sitter universe is
gone and our result is what one would expect to get in the Minkowski case.
We have the vacuum expectation value of one component of the stress-energy
tensor,
$$   <T_{\mu \nu}   > \propto {{\delta(v)}\over {u^3}} $$
where  $\delta(v)$ is the Dirac delta function.  In the Nutku-Penrose metric
the only non-zero component of the curvature tensor is proportional to the
same function.
One may anticipate, following Deser [1], that any non-zero result may be
proportional to the same form.

Our conclusion is that to obtain a non-zero result for the vacuum fluctuations
in the case studied here, first we have to work in de Sitter space, and thus
explicitly break the conformal symmetry in the metric we perturb about.
When this  calculation is carried to second order, infrared divergences are
traded for the  for the curvature of the space to get a finite result.
At the end we are back in the metric which describes an impulsive wave in
Minkowski space.

In section 2 we describe the second-order calculation in Minkowski space.
We pass to the de Sitter case just by multiplying the Minkowski result
by a conformal factor.  Before we go to the coincidence limit we set the two
auxiliary parameters proportional to each other.  We then take the
coincidence limit and perform differentiations on the Green function to
obtain a finite value for the vacuum expectation value of the stress-energy
tensor.  We conclude with a few remarks.
\bigskip
\noindent
{\bf{2. Minkowski calculation }}

We start with the Nutku-Penrose metric [3]
$$ds^2=2dudv-u^2|d\zeta+v\Theta(v0 f(\overline{\zeta}) d\overline{\zeta}|^2.
\eqno (1) $$
Here $v$ is the retarded time, $u$ is similar to the radial distance and
$\zeta$ is the angle of the stereographic projection, $f(\zeta)$ is the
Schwarzian derivative of  $h(\zeta)$  which is the holomorphic warp fuction
describing the impulsive wave, and $\Theta(v)$ is the Heavyside unit step
function.  For different choices of the warp function, the non-zero component
of the curvature tensor is multiplied by a different function, but the
essential characteristics  of the metric do not change.  The Nutku-Pentrose
solution corresponds to the snapping of a cosmic string, giving rise to a
spherical impuŸsive wave.

In the past we calculated vacuum fluctuations by computing the stress-energy
tensor of a scalar field in the background of this metric.  Our warp
functions  $h(\zeta)$  included a parameter which is related to the
string parameter  $G\ mu \approx 10^{-6} $ that we wrote as $\delta$ or
$\epsilon$.
We used $ h(\zeta) =\left(  {{ A \zeta+ B } \over {C\zeta + D}}
\right)^{1+\delta+i\epsilon}$        for
different values of (A,B,C,D) [4-8].
We got the null result in all of these cases if we perturbed around the
Minkowski space.  If we multiply this metric by a conformal factor,                          ,
$\left(1+{{\Lambda uv}\over {6}} \right)^{2} $ 
we get a spherical wave in de Sitter space [11].  Then there are finite
fluctuations even in first-order perturbation theory that are proportional to
the square of the scalar curvature [12].

To investigate the same phenomena in second-order  perturbation theory, here
we use a different choice of the warp function.
We take $h= e^{\alpha \zeta} $.
There are two reasons why this function is chosen.  First, this is an
important special case, first discussed by Gleiser and Pullin [13].
It has the special property that $f(\zeta)$ is independent of zeta although
$h(\zeta)$, the decisive term in the
solution, depends on  the same parameter.  This fact allows us to write
our solutions as sums of functions of $\zeta$ and $\overline {\zeta}$.
Our essential motive for studying this case is not this technical point,
though.  A similar problem was solved exactly [14], in another context.
There were somewhat vague indications of particle
production.  Here we want to see whether the vacuum expectation value of one
component of the stress-energy  tensor is proportional to the non-zero
component of hte curvature tensor.  Although
the result concerning particle production would not allow us to conclude
anything definite about our new calculation, we may still argue that an
unambigious result should exist for our latter calculation as well.
We will just use the exact result as a guard against possible infrared
problems.  If such problems arise in our expansion we will
know that they are due to technical factors, since the exact solution does
not have them, and we will try to regularize them.

We start with  $h= e^{\alpha \zeta} $ .  This gives $f=-{{\alpha^2}\over{2}}$
resulting in a metric
$$ds^2=2 dudv-{{1}\over{4}}\left(dx^2(2u-v\alpha^2 \Theta (v))^2
+dy^2(2+v\alpha^2 \Theta(v))^2\right). \eqno (2) $$
Here $\zeta=x+iy$.  If we write
the d'Alembertian operator in this metric , we get
$$ {{1}\over{\sqrt{-g}}} \partial_{u}(g^{\mu \nu} \sqrt{-g} ) \partial_v=
2\partial_{\mu}\partial_{\nu} + \left( {{1}\over {u-{{\alpha^2 v}\over {2}}} 
+{{1} \over {u+{{\alpha^2 v} \over {2}}}}} \right) \partial_v $$
$$+ {{\alpha^2}\over{2}} \left( {{1}\over{u+{{\alpha^2 v}\over {2}}}}
-{{1} \over {u-{{\alpha^2 v} \over {2}}}} \right) \partial_u
-{{1}\over {(u-{{\alpha^2 v}\over {2}})^2}} \partial_x^2 
-{{1}\over {(u+{{\alpha^2 v} \over {2}})^2}}
\partial_y^2. \eqno (3) $$ 
for the exact operator.
We multiply the d'Alembertian operator given in equation (3) by $u^2$ and
expand the operator up to second order in $\alpha^2$:
$$L^{II}=2u^2 \partial_{u} \partial_{v} +2u \partial_{v} -\partial_{x}^2
-\partial_{y}^2-{{\alpha^2v}\over{u}}(\partial_x^2-\partial_y^2)-\alpha^4
\left( {{v}\over{2}} \partial_u -{{v^2}\over{u}} \partial_v+{{3}\over {2}}
{{v^2}\over{u^2}}(\partial_x^2+\partial_y^2) \right). \eqno (4) $$
We can construct the vacuum expectation value of the stress-energy tensor
from the two-point function through differentiation, after the coincidence
limit is taken and  the appropriate regularization is done.  The two-point
function $G_F$ is found by summing the eigenfunctions of the related
Sturm-Liouville problem.

In zeroth order we get the empty space solution from the solution
$\phi^{(0)}$ [5]:
$$ G_F^{(0)}= \sum_{\lambda} {{\phi_{\lambda}^{*(0)}(x)
\phi_{\lambda}^{(0)}(x')} \over {\lambda}}=
{{A}\over{ (u-u')(v-v')-{{uu'}\over{2}}\left((x-x')^2+(y-y')^2 \right)}} .
\eqno (5) $$
Here $A$ is a constant.  The first order solution is written as 
$\phi^(1)=\phi^(0) f$.  Here $\phi^{(0)}$ is the zeroth-order solution.
The first order solution is the product of the zeroth-order solution and
another function.  This ansatz for $\phi^{(1)}$ is dictated by the
differential equations, and is not just an ad hoc guess.  
We find that $f$ just modulates the zeroth-order solution, and does not
essentially
change it.   The signature of $\phi^{(0)}$ is seen in the ultra-violet
behaviour of$G_F$ to a large extent.

We find that $f$ obeys the differential equation
$$L_2 f ={{ v}\over {u}} (k_2^2-k_1^2) \eqno (6) $$
where $L_2$ is defined as
$$L_2=\left( -2iR {{\partial} \over { \partial s}} 
-2i\left( k_1{{\partial} \over {\partial x}}
+k_2{{\partial}\over {\partial y}} \right) -{{\partial^2}\over {\partial x^2}}
-{{\partial^2}\over {\partial y^2}}-
2{{\partial ^2} \over {\partial s \partial v}}
+{{iK}\over {R}}{{\partial} \over {\partial v}} \right) \eqno (7) $$
Here  $k_1,k_2,R$ , and $K$ are parameters of the zeroth-order solution,
$s={{1}\over{u}}$.  They are integrated over to get the two-point function $G_G$.

To calculate $f$ we make the ansatz  $f=vf_1(s,x,y)+f_2(s,x,y)$ as
suggested by the form of equation (6).  This ansatz yields two
equations, partly coupled:
$$L_3f_1=s(k_2^2-k_1^2)  ,\eqno (8) $$
and
$$L_3 f_2=\left(2 {{\partial} \over {\partial s}}+{{iK}\over{R}} 
\right) f_1 \eqno (9) $$
where
$$L_3=\left( -2iR{{\partial}\over {\partial s}}-2i \left(
k_1{{\partial}\over{\partial x}}+ k_2 {{\partial}\over{\partial y}} \right)
-{{\partial ^2}\over {\partial x^2}}
-{{\partial ^2}\over{ \partial y^2}} \right). \eqno (10) $$
These equations are simply integrated over by multiplying the right-hand side
by the inverse of the operator $L_3$.

We end up with
$$f={{-ivs^2}\over{4R}} (k_1^2-k_2^2)+i(k_1^2-k_2^2)
\left({{is^2}\over {4R^2}}
+{{Ks^3}\over{24R^3}} \right). \eqno (11) $$
To get $G_F$, we have to calculate $O[f]$ where the operator $O$ is given by
$$O={{i}\over{(2\pi)^2uu'}} \int_{-\infty}^{\infty} {{dR}\over{2|R|}}
\int_{-\infty}^{\infty} dk_1 \int_{-\infty}^{\infty} dk_2 \int_{-\infty}
^{\infty} dK \int_{0}^{\infty} d\alpha $$
$$\times e^{ik_1(x-x')} e^{ik_2(y-y')} e^{ir(v-v')}
e^{{{iK}\over{2R}}(s-s')} e^{i\alpha (K-k_1^2-k_2^2)}. \eqno (12) $$
When we perform this operation, we get $G_F$ equal to
$$[(x-x')^2-(y-y')^2]\left(A_1{{s^2v\Theta(v)-{s'}^2v'\Theta(v')}
\over{(s-s')[\quad   ]}} \right. $$
$$ \left. +A_2{{\Theta(v)s^2+\Theta(v'){s'}^2}\over{(s-s')^2[\quad   ]}}
+A_3{{s^3\Theta(v)-{s'}^3 \Theta(v')}
\over {(s-s')^3[\quad   ] }} \right) \eqno (13) $$
where $ [\quad   ] = (u-u')(v-v')-{{uu'}\over {2}} 
\left( (x-x')^2+(y-y')^2 \right) $, 
and $A_1, A_2, A_3$ are constants.
We see that this result is of the Hadamard form.
No problems seem to arise in the infrared region. We find that all these
terms have the same type of ultraviolet singularity as the flat part.
We could not find a finite part of this expression.

If we go one order higher, we end up with the differential equation
$$L_2 g=v^2 \left( {{iRs}\over{2}}+3s^2(k_1^2+k_2^2)+(k_1^2-k_2^2)^2
\left({{-is^3}\over {4R}} \right)
\right) $$
$$+v\left( {{-s}\over{2}}+{{iKs^2}\over {4R}}-
{{(k_1^2-k_2^2)^2s^3}\over{4R^2}}
+{{iK (K_1^2-k_2^2)^2s^4}\over{24R^3}} \right) \eqno (14) $$
when we make the ansatz $\phi_2=\phi_0 g$.  We take
$g=v^2g_1(x,y,s)+vg_2(x,y,s)+g_3(x,y,s)$.
Going through similar steps as described above, we find
$$g_1= {{-s^2}\over{8}} +{{is^3(k_1^2+k_2^2)}
\over{2R}} +{{(k_1^2-k_2^2)^2 s^4}\over {32 R^2}}. \eqno (15) $$
All these terms give two-point functions,$G_F$, in the Hadamard form.
All are finite in the infrared sense.

We use $g_1$ to get $g_2$:
$$g_2= {{-i3s^2}\over {8R}}-{{13Ks^3 }\over {12R^2}}+{{is^4}\over{32 R^3}}
\left( (k_1^2-k_2^2)^2
 +K(k_1^2+k_2^2) \right) $$
$$ +{{s^5K}\over{160R^4}}(k_1^2-k_2^2)^2 .\eqno (16) $$
At this point we start getting problems.  All the terms in this expression
are divergent in the infrared region.  When we apply the operator $)$ to $g_2$
and take $x=x',y=y'$, we get a term proportional to
$$\int_{0}^{\infty} {{d\alpha}\over {\alpha}}
exp \left[{{-i(u-u')(v-v')}\over{uu'\alpha}} \right]. $$
We can start with a massive scalar particle and set the mass equal
to zero at the end of the   calculation.
Then the above integral reduces to $H_0(m\sqrt{(u-u')(v-v')} ) $
where $H_0$ is the Hankel function of order zero.  It degenerates to a
logarithm when $m$ tends to zero.  We have to take derivatives of $G_F$ to
find the vacuum  expectation value of the stress-energy tensor.  The term
with $m$ decouples if we differentiate the logarithm with respect to $u$
or $v$.  The finite part of $<T_{\mu \nu}>$ will not have $log m$
if there is a finite part.  We call this type of divergence 'harmless',
in this sense.

For$g_3$, we get
$$g_3= {{3s^2}\over {8R^2}}-{{i55Ks^3}\over {16R^3}} -{{s^4}\over{32 R^4}}
\left( (k_1^2-k_2^2)^2+K(k_1^2+k_2^2)+{{13K}\over{3}} \right) $$
$$+{{is^5K}\over{320R^5}}\left(3(k_1^2-k_2^2)^2+K(k_1^2+k_2^2) \right)
-{{K^2s^6(k_1^2-k_2^2)^2}\over{1920R^6}} .\eqno (17) $$

We see that when $x=x',y=y'$ all these terms give rise to expressions
that are proportional to
$$ \int_0^{\infty} d\alpha exp \left[ -i{{(u-u')(v-v')}
\over {uu' \alpha}} \right] \eqno (18) $$
which are linearly divergent at the upper limit .
If we use a massive field as an infrared cut-off, we get terms that go as 
${{1}\over {m^2}}$ as $m$ tends to zero.  This term multiplies the
whole expression and does not drop out on differentiation.
\bigskip
\noindent
{\bf{3.Going to de Sitter space}}
Up to this point we have studied a metric for an impulsive wave propagating
through minkowski space.  We ran into problems which indicate that we may be
perturbing around the wrong vacuum.  Since the exact treatment of a related
problem has no infrared divergences, we know that these divergences are
spurious  .

As a possible way out we take the impulsive waves propagating in a de Sitter
universe.  Since the de Sitter space has a parameter with the dimensions of
mass, one may think of trading thesec two parameters for one another.

we try to perturb around the de Sitter space.  Since the impulsive wave
solution in the de Sitter space [11] is just the Minkowski solution
multiplied by the factor $(1+{{\Lambda uv}\over {6}}$  , we get the
de Sitter two-point function just by multiplying the Minkowski case by
$(1+{{\Lambda uv}\over {6}} (1+{{\Lambda u'v'}\over {6}}) $.
This expression can be expanded as
$$\left( 1+{{\Lambda uv}\over {6}} \right) \left( 1+
{{\Lambda u'v'}\over {6}}\right) =
\left( 1+{{\Lambda UV}\over {6}} \right)^2+{{\Lambda}\over{12}}(u-u')(v-v')
+\Lambda^2   (...). \eqno(19)$$
Here $U={{u+u'}\over{2}},V={{v+v'}\over {2}}$.  We multiply the
Green function obtained in Minkowski space by this expression to get the
de Sitter expression.  In the Minkowski space $g_3$ is given above gives
rise to  $G_F$ whose first term goes as
$$ {{\Theta(v) s^2}\over {(s-s')^2m^2}}+{{\Theta (v'){s'}^2}\over
{ (s-s')^2m^2}}  \eqno (20) $$
for x=x',y=y'.  Here $m^2$ is an infrared parameter.  We introduced the
infrared  parameter by adding $2m^2u^2$ to our operator in equation (4).
When the calculation is done in the presence of this additional term, we get
the expressions given in equation (20) when we perform the summation operation
to obtain the propagator function.  In the presence of the new term the
operators given in expression (18) are modified and result in terms given in
(20).  When we go to de Sitter space we have to multiply them by the
expression given in (19).

Note that going to de Sitter space was only a technical trick.  We will
take $\Lambda$ equal to  zero at the end of the calculation and land in 
Minkowski space.   We see that many of the terms that part of Minkowski
space is subtracted for regularization are set to zero when $\Lambda$
goes to zero.  The terms given above are undeterminedsince they are
multiplied by ${{\Lambda}\over {m^2}}$, where both $m^2$ and $\Lambda$
tend to zero.  At this point we choose $\Lambda$ proportional  to $m^2$.
This choice is dimensionally correct.  Any other choice, say, $\Lambda$
proportional to the first power of$m$ times $s$, would be unnatural.
To obtain the vacuum expectation value of the stress-energy tensor, we have
to differentiate the Green function with respect to the coordinates.
We are particularly interested in $<T_{\mu \nu}>$.  We first take $m$
and $\Lambda$ going to zero limit, and then differentiate with resoect
to $v$ and $v'$ symmetrically, and then take the coincidence limit.
We see that at the end of this calculation $<T_{\mu \nu}>$ turns out
to be proportional to ${{\delta(v)}\over {u^3}}$.
Here $\delta$ is the Dirac delta function.  The proportionality constant
between the scalar curvature and the infrared mass squared can be absorbed
in the   perturbation constant $\alpha ^2$.
\bigskip
\noindent
{\bf{Conclusion}}
\noindent
We tried to calculate here the quantum fluctuations resulting from snapped
cosmic strings, by perturbing around the vacuum, for a warp function that
corresponds to the Gleiser-Pullin solution [14].  we could not seperate a
finite part in the first-order calculation.  in the second order, we ran
into $infrared$ divergences.  If we use the same warp function in the
calculation where we perturb around the de Sitter space and take the
scalar curvature $\Lambda$  proportional to $m^2$, we can get finite
results in second-order perturbation theory.
In the above expression $m$ is the infrared cut-off.  At the end we get
$<T_{\mu \nu}>$ proportional to ${{\delta(v)} \over {u^3}}$.  This is just the
result we anticipated in the Minkowski case.  At the end we are back in
Minkowski space.

This result suggests that to get finite results in perturbation theory for the
case studied we have first to break the conformal invariance which does not
allow vacuum fluctuations for the massless particles [10], by hand, by
going to de Sitter space, and then come back to Minkowski space after the de
Sitter space parameter cancels the infrared divergence.  As a result of
this operation we get a non-zero contribution for vacuum fluctuations.

We anticipate somehow detecting the presence of cosmic strings [15].
We were expecting to get a finite vacuum expectation value for the
stress-energy.  We could not get finite contributions in our previous
calculations when we stayed in Minkowski space [4-7].  When similar
calculations were performed for the de Sitter case, we got finite
results which were proportional to positive powers of the curvature of the
de Sitter space. [12,8].  The contributions were zero when the curvature
was set to zero.

In second-order calculations we encountered infrared divergences, which made
the expansion ambigious at this order.  We cam give a meaning to these
calculations  by first starting with the de Sitter impulsive wave where the
curvature of the space is proportional to the square of the mass of the scalar
field whose stress-energy is calculated in the background metric of the
impulsive wave.  This mass is used as an infrared parameter further in the
calculation.

Our method may be criticized  because it was applied only to a single choice
for the warp function.  although our choice was a very important case,
we tried to perform the same calculation for the case where
$$ h(\zeta)= \left({{1}\over {\zeta}}\right)^{1+i\epsilon}  \eqno (21) $$
and expanded the operator in powers of $\epsilon$.  This choice is
known to behave exactly the same way as the more general case,
$$h= \left( {{A \zeta+B} \over {C \zeta+D}} \right)^{1+\delta+i\epsilon}
 \eqno (22) $$
the second-order equation reads
$$ (L_0-\lambda_0) \phi^{(0)} = -L'_2 \phi^{(0)}-L'_1 \phi^{(1)}
 \eqno (23) $$
where
$$L'_2={{8v}\over {u}} \left( v{{\partial} \over {\partial v}}
-u{{\partial} \over {\partial u}} \right)
+F[v,u,x,y,{{\partial}\over {\partial x}}, {{\partial}\over {\partial y}}].
\eqno (24) $$
Here $\phi_0,\phi_1$ and $\phi_2$ are the zeroth-,first- and second-order
solutions.  The first term on the right-hand side of this expansion gives
terms that read as
$${{-8v}\over{u}} \left[ iRv+1-{{iK}\over {2uR}} \right] \eqno (25) $$
which are of the same form as the terms in equation (16).  When these terms
are integrated, we obtain
$$ \phi^{(2)}= \phi^{(0)} \left( 2{{v^2}\over {u^2}} +{{2iv}\over {Ru^2}}
-{{1} \over {R^2u^2}}-{{iK}\over {3R^3u^3}} \right)+... \eqno (26) $$
where dots represent the additional terms we get from those included in F
in equation (24).

Doing the same calculation to obtain Green function, we get an expression
that behaves exactly as those given in equation (20).  There are no terms
which cancel these terms.  At the end $<T_{\mu \nu}> $ comes out to be
proportional to ${{\delta(v)}\over {u^3}}$.  The details of this
calculation will  be published elsewhere.

We think that this second calculation shows that our result is not restricted
to one particular case, but illustrates the general behaviour in this
problem.  By starting with a massive field in the background metric of the
impulsive wave in de Sitter space, we are breaking conformal invariance
by hand.  In the absence of the impulsive wave, i.e. in the zeroth-order
perturbation theory, we do not have conformal invariance in the model,
in contrast   to the case when we have a massless scalar field and an
impulsive wave in the minkowski space.  Then we set the two parameters
proportional to each other and send them to zero.
As a result of this operation we get a finite contribution for
$<T_{\mu \nu}>$     .

Our problem did not have conformal invariance in the first place.
Using perturbative methods we could not convey this information to our
solutions.  We propose   , then, first introducing additional parameters
where this invariance is broken by hand in the perturbative calculation.
At the end we send these additional parameters to zero in thwe same
manner.  We found that the peerturbative calculation gave the same
qualitative    answer, at least, that an exact solution would have given.
\bigskip
\noindent
{\bf{Acknowledgments}}
\noindent
I thank Professor Y. Nutku and Professor A. N. aliev for many fruitful
discussions of this problem.  I am also grateful to Professor Nutku for
a critical reading of the manuscript.  I thank Dr. N. \" Ozdemir for
discussions and B. Yap\' i\c skan  for checking first-order calculations.
This work is partially supported by T\" UBITAK, the Scientific and
Technical Research Council of Turkey and the Academy of Sciences of Turkey.
\bigskip
\noindent
{\bf References}

\item{1.} S.Deser, J. Phys. A: Math. Gen. {\bf{81}} (1975) 972;

\item{2.} G.W.Gibbons, Commun. Math. Phys. {\bf{45}} (1975) 191;

\item{3.} Y.Nutku and R. Penrose, Twistor Newslett. {\bf{34}} (1992) 9;

\item{4.} M. Horta\c csu, Class. Quant. Grav. {\bf{7}}(1990) L165,
{\bf{9}} (1992) 799;

\item{5.} M.Horta\c csu , J. Math. Phys. {\bf{34}}(1993) 690;

\item{6.} M.Horta\c csu and R. Kaya, J. Math. Phys. {\bf{35}} (1994) 3043;

\item{7.} M.Horta\c csu, R. Kaya and N. \" Ozdemir, {\it{ Lecture Notes
in Physics: Strings and Symmetries}}, ed. by G. Akta\c s, C. Sa\c cl\i o\u glu
and M. Serdaro\u glu (Berlin; Springer) pp. 297-302;

\item{8.}  Y. Enginer, M.Horta\c csu and N. \" Ozdemir, 1995 TUBITAK MAM
preprint;

\item{9.}  A.N. Aliev, private communication;

\item{10.}  N.D. Birrell and P.C.W. Davies,
{\it{Quantum Fields in Curved Space}} (Cambridge: Cambridge University Press)
1982;

\item{11.}  P.A.Hogan, Phys. Lett. {\bf{171A}} (1992) 21;

\item{12.} M.Horta\c csu and N. \" Ozdemir  , 1995   ITU report;

\item{13.} R.Gleiser and J. Pullin, Class. Quant. Grav. {\bf{6}} (1989) L141;

\item{14.}  A.H. Bilge, M. Horta\c csu and N. \" Ozdemir, Gen. Rel. Grav.
{\bf{28}} (1996) 511;

\item{15.} B.F.Svaiter and N.F.Svaiter, Class. Quantum Grav. {\bf{11}}
(1994) 347, J. Audretsch, R. Muller and M. Holzman, Class. Quant. Grav.
{\bf{12}} (1995) 2927, L. Iliadakis, V. Jasper and J. Audretsch, Phys. Rev.
D {\bf{51}}(1995) 2591.
\end